**Variable selection in social-environmental data: Sparse regression and tree ensemble machine learning approaches**

Elizabeth Handorf†[1], Yinuo Yin[2], Michael Slifker[1], and Shannon Lynch‡[2]

[1]Biostatistics and Bioinformatics Facility, Fox Chase Cancer Center, Philadelphia, PA, USA

[2]Cancer Prevention and Control, Fox Chase Cancer Center, Philadelphia, PA, USA

†Email: Elizabeth.Handorf@fccc.edu

Phone: 215-728-4773

Reimann 383

333 Cottman Ave

Philadelphia, PA 19111, USA

‡Email: Shannon.Lynch@fccc.edu

Phone: 215-728-5377

Young Pavilion

333 Cottman Ave

Philadelphia, PA 19111, USA

*Keywords:* Variable selection; Social environment.

---

† Contributions equal for senior and corresponding authors

**Abstract**

**Objective:** Social-environmental data obtained from the U.S. Census is an important resource for understanding health disparities, but rarely is the full dataset utilized for analysis. A barrier to incorporating the full data is a lack of solid recommendations for variable selection, with researchers often hand-selecting a few variables. Thus, we evaluated the ability of empirical machine learning approaches to identify social-environmental factors having a true association with a health outcome.

**Materials and Methods:** We compared several popular machine learning methods, including penalized regressions (e.g. lasso, elastic net), and tree ensemble methods. Via simulation, we assessed the methods' ability to identify census variables truly associated with binary and continuous outcomes while minimizing false positive results (10 true associations, 1,000 total variables). We applied the most promising method to the full census data (p=14,663 variables) linked to prostate cancer registry data (n=76,186 cases) to identify social-environmental factors associated with advanced prostate cancer.

**Results:** In simulations, we found that elastic net identified many true-positive variables, while lasso provided good control of false positives. Using a combined measure of accuracy, hierarchical clustering based on Spearman's correlation with sparse group lasso regression performed the best overall. Bayesian Adaptive Regression Trees outperformed other tree ensemble methods, but not the sparse group lasso. In the full dataset, the sparse group lasso successfully identified a subset of variables, three of which replicated earlier findings.

**Discussion:** This analysis demonstrated the potential of empirical machine learning approaches to identify a small subset of census variables having a true association with the outcome, and that replicate across empiric methods.

**Conclusion:** Sparse clustered regression models performed best, as they identified many true positive variables while controlling false positive discoveries.

## 1. BACKGROUND AND SIGNIFICANCE

The Precision Medicine Initiative suggests that environment, along with genes and lifestyle behaviors, should be considered for cancer treatment and prevention. Nevertheless, the impact of social environment, or the neighborhood in which a person lives, remains understudied. (1) Compared to the biological level where empirical, high-dimensional computing approaches, like genome-wide association studies (GWAS), are often used for hypothesis-generation, risk prediction, and variable selection, empirical methods are only beginning to be employed at the environmental level. (2, 3)

Social environment, as defined by a patient's neighborhood of residence, is particularly relevant to the study of cancer health disparities. Neighborhood boundaries can be defined by U.S. Census tracts (smaller geographic areas than a county). These neighborhoods can be described by variables measuring economic (e.g., employment, income); physical (e.g., housing/transportation structure); and social (e.g., poverty, education) characteristics. (4, 5) Studies linking U.S. Census data with state and national cancer registry data show that neighborhood can help explain differential cancer incidence and mortality rates beyond race/ethnicity or genetic ancestry, and that neighborhood environment often exerts independent effects on cancer outcomes. (6, 7)

Methodological challenges have limited the incorporation of neighborhood data into Precision Medicine. Most studies use *a priori* variable selection approaches, but there are no standard variables to represent particular domains (e.g. poverty, education, employment, etc.), which has limited translation of social environmental variables into clinical use. In the few studies using empiric selection approaches, variable selection and replication of findings were complicated by the high degree of correlation among U.S. Census variables. For instance, similar to a GWAS, we previously conducted a neighborhood-wide association study (NWAS) in both black and white men in Pennsylvania and agnostically identified 22 U.S. census variables (out of over 14,000) significantly associated with advanced prostate cancer. (3) In the first NWAS, social support was identified as an important neighborhood domain, but 2 very similar variables were identified to represent this domain *(% male householders living alone* vs *%male householders over 65 living alone in a non-family household)*. Thus, multicollinearity (the presence of many highly interrelated variables) is a challenge for variable selection and replication.

The systematic assessments offered by machine learning algorithms, which allow for high dimensionality and collinearity, may be useful for analyses of neighborhood data. In this manuscript, we broadly use the term "machine learning" to refer to any computational method which selects variables automatically, without direct input from a human analyst. While the main objective of machine learners is often predictive accuracy, an additional objective is variable selection and determining which features are truly important. This is analogous to the goals of variant discovery vs risk prediction in genetic studies. (8, 9)

Motivated by the previous NWAS of prostate cancer cases in Pennsylvania, we sought to understand which machine learning algorithms are most effective for identification of neighborhood factors which have a true association with a health outcome. Machine learning algorithms are often judged by comparing predicted vs. observed outcomes in an independent test set. We cannot use this paradigm to evaluate variable selection, however, as the true underlying variables associated with a given outcome are unknown. This motivated use of a simulation study, where we generated outcomes that are dependent on a small subset of the potential predictor variables. We then applied several popular machine learning approaches, including lasso, elastic net, hierarchical clustering, and random forests, and assessed how well each method identified true positive variables while minimizing false negatives. We compared the results to traditional regression with correction for multiple testing. Finally, we applied the top performing machine learning approaches to our original NWAS dataset, and compared findings from these analyses to our first NWAS in white men.

**MATERIALS AND METHODS**

**1.1.  Candidate Methods for discovery of important variables**

Below, we describe methods for variable selection where both $p$, the number of potential predictor variables, and $N$, the number of observations, are large, and discuss how these methods can be applied to analysis of neighborhood-level covariates. We identified methods which provide objective and automatic variable (feature) selection for both continuous and binary outcomes.  We also limited our evaluation to methods with largely automated tuning, which are readily implemented using standard R packages, and which one can run within reasonable timeframes. Ultimately, the methods we identified fell within two broad categories: penalized models, and ensemble tree-based methods.

### 1.1.1. Standard regression models

The simplest approach to variable selection is similar to the GWAS and NWAS approach. (10) A series of univariable tests are conducted to determine the relationship between each possible predictor and the outcome. Variables which are statistically significant after correction for multiple-testing (11) (i.e. 'top hits') are then replicated in a separate set of samples. (12) Although this approach is simple and easy to implement, the separate regression models ignore any correlation structure between candidate predictors. This may lead to selection of a large number of highly correlated variables, necessitating further variable selection steps, as described in the NWAS manuscript. (10) We included this approach as a baseline for comparison, to demonstrate the degree of improvement more advanced methods can provide.

### 1.1.2. Sparse regression models

Penalized regression addresses some of the limitations of standard regression for high-dimensional data. A useful class of these models provide shrinkage which enforces sparsity; that is, many of the parameter estimates are shrunk to exactly zero. (13) Sparse models have several advantages over traditional regression, including reduced overfitting (which improves prediction), accommodating multicollinearity, and the ability to fit models where $p > n$. They can also be used for variable reduction, where a zero parameter estimate indicates that the variable is not an important predictor.

#### 1.1.2.1. Lasso penalized regression

The Least Absolute Shrinkage and Selection Operator (lasso) includes a L1-norm (absolute value) penalty that shrinks many parameter estimates to exactly zero. (13, 14) Thus, variables with non-zero coefficients can be considered the important predictors for the outcome of interest.

For a linear regression, the lasso solution is found as

$$\min_{\beta} \left\{ \frac{1}{2N} \sum_{i=1}^{N} \left( y_i - \sum_{j=1}^{p} x_{ij} \beta_j \right)^2 + \lambda \sum_{j=1}^{p} |\beta_j| \right\}, \qquad (1)$$

where $N$ is the number of observations, $p$ is the number of parameters, $Y$ is a vector of outcomes, $X$ is a $N$ x $p$ matrix of covariates, and $\beta$ is a vector of effects. The size of the penalty is determined by $\lambda$, which can be found via cross-validation to minimize prediction error. Alternatively, one can choose a stricter threshold for $\lambda$ at 1 standard error above the minimum prediction error (to conservatively allow for error in the estimate of the optimal $\lambda$). (13) Although the lasso can find a solution under multicollinearity, if a group of highly correlated variables is present the lasso tends to arbitrarily select one variable and set the others to zero. (15)

*1.1.2.2. Elastic net*

The elastic net was proposed to overcome some of the limitations of the lasso method. It uses a combination of the L1 lasso penalty and the L2 ridge penalty:

$$\min_{\beta}\left\{\frac{1}{2N}\sum_{i=1}^{N}\left(y_i - \sum_{j=1}^{p}x_{ij}\beta_j\right)^2 + \lambda\left(\frac{1}{2}(1-\alpha)\sum_{j=1}^{p}\beta_j^2 + \alpha\sum_{j=1}^{p}|\beta_j|\right)\right\}, \qquad (2)$$

where $0 \leq \alpha \leq 1$, and other parameters are defined as above in (1). The choice of $\alpha$ determines whether the penalty is closer to a ridge penalty ($\alpha=0$) or a lasso penalty ($\alpha=1$). The choice of both $\alpha$ and $\lambda$ can be determined via cross validation. (15) Unlike the lasso, when predictors are collinear, the elastic net tends to classify groups of highly correlated variables as all either zero or nonzero. In many cases the elastic net provides better performance than the lasso. (15)

*1.1.2.3. Sparse models with clustering*

Hierarchical clustering is a way of grouping variables with similar behavior across observations. Agglomerative clusters are built from the bottom-up by joining the "closest" clusters at each step according to defined distance and linkage functions, and the distance becomes the "height" at which clusters are joined. For the census data we propose to define distance as one minus the absolute value of the Spearman correlation coefficient. Complete linkage is a useful choice here as it maintains the original scale of the distance measures (in this case, from 0 to 1), and the height is therefore interpretable. The resulting clusters are represented via a dendrogram (see Additional file 1). (16)

Cluster membership can be defined by cutting the dendrogram at a specific height, so that any observations that are joined at a height lower than that value are members of a cluster. A more objective method is to identify statistically significant clusters via a bootstrap. (17) This method resamples participants to identify which clusters of variables often appear, measuring stability. Note that with either method, many clusters may contain only one variable. If the number of clusters is small relative to *n*, clustered variables can be summarized into a single measure, and models can then be fit using multiple regression models. However, if a large number of clusters are present, a better choice is to use cluster membership to fit group lasso or sparse group lasso models. The sparse group lasso is particularly useful, as it has penalties at both the group and individual level, allowing for sparsity across and within groups. (18-20)

### 1.1.3.    Tree ensemble methods

Another group of popular machine learning methods are based on tree ensembles, where many decision trees are fit to the data.  Decision trees rely on recursive binary partitioning, that is, at each step (node) in the tree, the observations are split into two daughter nodes depending on some function of the predictor variables.  Often, methods aggregating many trees (ensembles) outperform single tree based methods. (21)

#### 1.1.3.1. Random forests

The random forests method is useful for high-dimensional data.  Underlying the method are many Classification and Regression Trees run on bootstrap samples of the dataset. (16) The relative impact of each variable on prediction accuracy is characterized using the variable importance score (VIMP), calculated by permuting each variable and re-fitting the random forest, and assessing how this impacts accuracy. VIMP scores provide a way of ranking predictors relative to each other, but choosing a threshold for the VIMP is often done post-hoc.

Recently, Ishwaran and Lu (2019) (22) proposed a resampling based calculation of the standard errors for the VIMP. We propose to use this standard error to inform variable selection via the confidence interval. Based on the estimated VIMP scores and their standard errors, we can create $100*(1-\alpha/p)$% confidence intervals for each variable. If the confidence interval excludes zero, we can conclude that there is evidence that the variable improves prediction, and therefore infer that there is a relationship between that variable and the outcome of interest.

#### 1.1.3.2 Bagging

Bagging was a predecessor to random forests, and can be thought of as a special case.  In the standard interpretation, at each node a random subset of variables (often choses to be $p/3$) are evaluated as candidates for splitting.  In bagging, all $p$ variables are considered for possible splitting.  This tends to yield a smaller subset of variables with high VIMP scores, which may be better for our purposes of identifying the best variables. (21)

We note that, as this is a special case of random forests, we can use the same resampling-based approach to define confidence intervals for the VIMP scores and accomplish variable selection.

### 1.1.3.3 *Bayesian Additive Regression Trees (BART)*

Like Random Forests and Bagging, BART is a tree ensemble method; however, it builds a set of trees using repeated draws from a Bayesian probability model. Similar to the VIMP of random forests, the relative importance of a given variable can be characterized using the variable inclusion proportion, defined as the number of splitting rules based on the variable out of the total number of splits. We can obtain an empirical estimate of the null distribution for the variable inclusion proportions by permuting the outcomes and re-fitting the BART algorithm. After these are obtained, three thresholds for variable selection have been proposed. The first, the local threshold, uses the null distribution of each individual variable, and if the fitted inclusion proportion is greater than its 1-α quantile under the null, that variable is selected. A more restrictive criterion (Global SE) increases the threshold using the local mean and standard deviation with a global multiplier determined based on the permutation distribution of all variables. The most restrictive criterion (Global Max) requires that the inclusion proportion is greater than the 1-α quantile of the maximum inclusion proportions (across all variables) from each permutation.

### 1.2. Simulation study

Machine learning methods are typically evaluated by their ability to predict outcomes. In this study, we are interested in a different question: how well does each method correctly identify the subset of census variables which are truly associated with the outcome of interest? Therefore, we conducted a simulation-based experiment, generating outcomes which have known associations with a small subset of census variables.

The data structure of the census variables is complex; measures may exhibit non-normal distributions, contain excess numbers of zeros, and some variables are highly collinear (see Additional File 2). To fully reflect this complexity, we used observed census variables for PA prostate cancer cases (10) as the basis of our simulation. For computational tractability, we randomly selected 1,000 variables and 2,000 individuals. Missing values were imputed using median substitution, and all variables were standardized (mean = 0, standard deviation = 1). Let $X$ be the data matrix corresponding to the full set of 1,000 neighborhood variables. We define $X_T$ as the matrix with columns representing variables truly associated with the outcome of interest, $Y$. The full set of predictors, $p$, also contains many other predictors. We define matrix $\boldsymbol{X_0}$ to be the matrix of columns not directly related to $Y$. The variables in $\boldsymbol{X_0}$ which are highly correlated with variables in $\boldsymbol{X_T}$ are considered "surrogate" variables.

We selected 10 variables to be members of $\boldsymbol{X_T}$, where 5 variables exhibited marked collinearity ($X_1$-$X_5$, correlation >0.95 with at least 1 other variable), and 5 variables which exhibited modest or low collinearity ($X_6$-$X_{10}$, correlation < 0.6 with all other variables). Outcomes were simulated according to the structure shown in Figure 1. We considered both binary and continuous outcomes as they are commonly seen in medical research with the mean models $\text{logit}(E(Y)) = \beta'\boldsymbol{X_T}$ for binary $Y$ and $E(Y) = \beta'\boldsymbol{X_T}$ for continuous $Y$, with errors distributed as $N(0,1)$. Effect size ($\beta$) was equal for each member of $\boldsymbol{X_T}$, with $\beta = 0.22$ for binary outcomes and $\beta = 0.11$ for continuous outcomes. The size of $\beta$ was set as the effect size in a single univariable test for which we would have at least 80% power with $5*10^{-5}$ 2-sided type-I error to detect the effect when $N$=2,000.

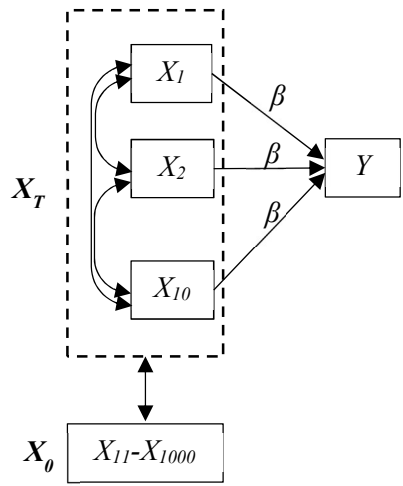

Fig. 1. Simulation model, showing the relationship between the ten elements of $\boldsymbol{X_T}$, the 990 elements of $\boldsymbol{X_0}$, and outcomes $Y$. Components of $\boldsymbol{X_T}$ and $\boldsymbol{X_0}$ may be correlated, denoted by double-sided arrows.

Table 1. Candidate methods

| Abbreviation | Description | Selection rule | R packages |
|---|---|---|---|
| UNIV-BFN | Univariable models with Bonferroni-adjusted p-val | $P < 5*10^{-5}$ | base R (23) |
| LASSO-MIN | Lasso with $\lambda$ chosen at the minimum prediction error | $\beta \neq 0$ | `glmnet` (24) |
| LASSO-1SE | Lasso with $\lambda$ chosen at 1 SE above the minimum error | $\beta \neq 0$ | `glmnet` |
| ELNET-MIN | Elastic net, grid search for $\alpha$ (0.05-0.95 by 0.05), $\lambda$ at min | $\beta \neq 0$ | `glmnet` |
| ELNET-1SE | Elastic net, grid search $\alpha$ (0.05-0.95 by 0.05), $\lambda$ at 1 SE | $\beta \neq 0$ | `glmnet` |
| HCLST-CORR-SGL | Hierarchical clustering, groups with corr > 0.8, sparse group lasso | $\beta \neq 0$ | `SGL` (25) |
| HCLST-BOOT-SGL | Hierarchical clustering, groups from bootstrap, sparse group lasso | $\beta \neq 0$ | `SGL, pvclust` (17) |
| RF | Random Forests algorithm with bootstrap-based confidence intervals for the variable importance scores | 99.995% CI > 0 | `randomForestSRC` (26) |
| BAGGING | Similar to Random Forests, but with all variables considered candidates for splitting at each node | 99.995% CI > 0 | `randomForestSRC` |
| BART-LOCAL | Bayesian Additive Regression Trees, local criteria for Inclusion Proportion (IP) | IP > 0.95 quantile of local distribution | `bartMachine` (27) |
| BART-GLOBALSE | Bayesian Additive Regression Trees, global SE criteria for IP | IP > threshold from local distribution with global multiplier | `bartMachine` |
| BART-GLOBALMAX | Bayesian Additive Regression Trees, global Max criteria for IP | IP > 0.95 quantile of global max distribution | `bartMachine` |

We simulated outcomes ($Y$) five-hundred times. In practice, variable selections are often internally validated by withholding a portion of the dataset. Therefore, we randomly assigned 2/3 of the data to be the discovery set and the remaining 1/3 to be the validation set. The algorithms discussed above were applied to each set of simulated outcomes. Candidate variables selected in the discovery set were then validated in the withheld 1/3 sample, using a series of univariable regression models, considering any variable with a P-val<0.05 to be validated, similar to the approach taken in some GWAS studies. (28) For the random forest method, which often identified a large number of variables in the discovery set, we also explored using a multivariable lasso model in the validation in an attempt to reduce potential confounding in the validation step. Table 1 lists each method tested, along with the selection rule in the discovery set. All models were fit using R software (version 3.5). (23) Software used to fit models and run simulations is available on github. (https://github.com/BethHandorf/neighborhood-machine-learning)

1.2.1.   Comparison of Methods: Performance Assessment

Performance was quantified by which methods identified a large proportion of true positive variables ($X_T$) and minimized false positive variables ($X_0$). We also considered more flexible success metrics where true positives were considered as the identification of either a target variable or a good surrogate (correlation > 0.8 with target), and false positives were considered as those not a target variable or a surrogate. The threshold of 0.8 was chosen as it is a commonly used value for determining suitability of surrogate endpoints in clinical studies. (29, 30) We also considered a composite metric, the F2 score.  This is a measure of accuracy combining the Positive Predictive Value (PPV, sometimes termed precision) and the sensitivity (sometimes termed recall). The F2 score is a specific case of the general F score, which gives greater weight to the sensitivity. (31)

$$F2 = \frac{(2^2 + 1) * TP}{(2^2 + 1) * TP + 2^2 * FN + FP}$$

Where TP is the number of true positives, FN is the number of false negatives, and FP is the number of false positives.  When comparing models, a F2 score closer to 1 denotes the preferred model. We chose the weighted F2 as we believe that in this application, priority should be given to the ability to detect more true positive variables. Finally, we determined the average detection rate for the individual variables, and evaluated how effect size estimates from univariable models were related to the likelihood of detection.

**1.3.   Application to PA prostate cancer cases**

To illustrate these methods in practice, we applied the most promising algorithms (those with the most true positives and fewest false positives) from the simulation study to the full dataset used in the prior NWAS study, which linked prostate cases from a PA Department of Health registry to social-environmental variables obtained from the U.S. Census (10) The binary outcome of interest was aggressive (high stage AND grade) prostate cancer(10).  This cohort of white prostate cancer patients diagnosed between 1995 and 2005 contained 76,186 individuals (8% with aggressive disease). There were 14,663 census variables evaluated for association with the outcome.  We included census variables as predictors, along with age and year of diagnosis. The data was split into discovery (2/3) and validation (1/3) samples. As above, variables selected in the discovery set were tested using univariable regression in the validation set, using a p-value cutoff of 0.05. We then compared which variables were selected by the most promising methods (from the simulation study) in the full study population to those found by the original NWAS method.

## 2.  RESULTS

### 2.1.  Comparison of methods

Table 2 shows the mean number of variables detected by each method, broken down into true positives and false positives. The strict definition considers true positives to be identification of variables in $X_T$, while the relaxed definition allows for surrogate variables. For the false positives, the strict definition shows the number of members of $X_0$ which were identified, while the relaxed definition shows the number of selected members of $X_0$ which did not have a substantial correlation with an element of $X_T$. The number of false positives was substantially reduced under the relaxed definition, especially for methods which identify groups of correlated predictors (e.g. elastic net, sparse group lasso), demonstrating that many of the "false positive" results were identified due to their relationship with a "true positive" variable.

For binary outcomes, the sparse regression method identifying the fewest false positive results was the lasso with the restrictive 1SE penalty (LASSO-1SE), while elastic net with the less restrictive penalty (ELNET-MIN) identified the largest number of true positives. When considering the sparse group lasso, the simpler correlation-based threshold to define clusters worked somewhat better than the more complex bootstrap-based cluster detection (although results were largely similar). The correlation-based clustering generated more clusters on average than bootstrap-based cluster selection (837 vs 772), so more restrictive cluster definitions may have led to these differences. Of the tree ensemble methods, BART-LOCAL performed the best.  It substantially outperformed both RF or BAGGING.  BART-GLOBALSE and BART-GLOBALMAX were too restrictive, identifying very few true positive variables.

Comparing the methods, there was generally a trade-off between the number of true positives and false positives. However, certain strategies were clearly inferior to others (dominated), with higher false positive rates and lower true positive rates than other methods. Univariable models and the random forests based models can be eliminated from consideration in future studies based on this criterion. When assessing the combined F2 measure of performance, many of the penalized models performed particularly well, with HCLST-CORR-SGL doing the best overall. BART-LOCAL also did well, particularly under the relaxed definition. The F2 measure indicates that these methods have particularly good sensitivity, while also considering their PPV.

Table 2: Simulation study results

| A. Binary outcome | Strict | | | Relaxed* | | |
|---|---|---|---|---|---|---|
| | TP (N/10) | FP (N/990) | F2 | TP (N/10) | FP (N/953) | F2 |
| UNIV-BFN | 4.09 | 32.49 | 0.267 | 5.13 | 12.70 | 0.443 |
| LASSO-1SE | 3.84 | 6.05 | 0.383 | 5.53 | 3.71 | 0.559 |
| LASSO-MIN | 4.25 | 9.01 | 0.399 | 5.98 | 6.53 | 0.569 |
| ELNET-1SE | 5.26 | 20.51 | 0.405 | 6.21 | 9.33 | 0.560 |
| ELNET-MIN | 5.53 | 26.11 | 0.393 | 6.61 | 14.40 | 0.548 |
| HCLST-CORR-SGL | **5.40** | **17.91** | **0.428** | **6.39** | **7.09** | **0.597** |
| HCLST-BOOT-SGL | 5.20 | 16.66 | 0.420 | 6.35 | 7.07 | 0.594 |
| RF | 3.53 | 18.41 | 0.281 | 4.91 | 7.68 | 0.462 |
| BAGGING | 3.56 | 13.94 | 0.308 | 4.73 | 6.70 | 0.456 |
| BART-LOCAL | 4.68 | 15.66 | 0.387 | 6.32 | 7.13 | 0.591 |
| BART-GLOBALSE | 1.96 | 0.53 | 0.228 | 2.24 | 0.22 | 0.261 |
| BART-GLOBALMAX | 0.01 | 0.00 | 0.001 | 0.01 | 0.00 | 0.001 |
| B. Continuous outcome | Strict | | | Relaxed* | | |
| | TP (N/10) | FP (N/990) | F2 | TP (N/10) | FP (N/953) | F2 |
| UNIV-BFN | 4.83 | 39.57 | 0.286 | 5.90 | 17.47 | 0.468 |
| LASSO-1SE | 2.88 | 4.49 | 0.298 | 4.33 | 2.42 | 0.454 |
| LASSO-MIN | 4.61 | 10.52 | 0.419 | 6.47 | 7.60 | 0.599 |
| ELNET-1SE | 3.88 | 8.27 | 0.366 | 4.87 | 3.29 | 0.500 |
| ELNET-MIN | 5.18 | 14.79 | 0.433 | 6.61 | 8.88 | 0.598 |
| HCLST-CORR-SGL | **5.82** | **19.86** | **0.444** | **6.81** | **8.51** | **0.617** |
| HCLST-BOOT-SGL | 5.52 | 17.03 | 0.441 | 6.72 | 8.45 | 0.610 |
| RF | 4.63 | 28.46 | 0.316 | 5.93 | 14.26 | 0.493 |
| BAGGING | 4.40 | 25.73 | 0.314 | 5.81 | 12.96 | 0.494 |
| BART-LOCAL | 5.14 | 18.35 | 0.404 | **6.80** | **8.34** | **0.617** |
| BART-GLOBALSE | 2.40 | 0.93 | 0.274 | 2.85 | 0.41 | 0.326 |
| BART-GLOBALMAX | 0.01 | 0.00 | 0.002 | 0.01 | 0.00 | 0.002 |

TP=True positive; FP=False positive; boldface denotes best performing method by F2 statistic
*Under the relaxed definition, if a true variable or its surrogate was selected, that variable was considered to be identified. Surrogates are therefore no longer in the pool of potential false positives, but the maximum number of true positive variables remains 10.

The results were largely similar for both continuous (Normal) and binary outcomes. We did find that Random

Forests (RF) identified more variables (both false positives and false negatives) for the continuous outcome,

while elastic net identified more variables with the binary outcome; however, the same method (HCLST-CORR-

SGL) had the highest F2 statistics for the strict and relaxed definitions. For the continuous outcome, under the relaxed definition, BART+LOCAL did as well as HCLST-CORR-SGL.

Considering the individual variables, those chosen from areas of high correlation ($X_1$-$X_5$) were selected less often than those from areas of moderate to low correlation ($X_6$-$X_{10}$), and there was more variability in detection rates for $X_1$-$X_5$. (See Additional File 3) Unsurprisingly, the lasso had notably low detection rates for $X_1$-$X_5$.

**2.2.  Performance Assessment:  Exploration of false negatives/impact of correlated data**

Table 3: Effect of confounding on detection rate (binary outcome)

|  | Mean β (true=0.22) | Detection rate (UNIV-BFN) | Detection rate (HCLST-CORR-SGL) | Detection rate RF with lasso validation |
|---|---|---|---|---|
| $X_1$ | 0.223 | 0.25 | 0.586 | 0.244 |
| $X_2$ | 0.343 | 0.972 | 0.778 | 0.126 |
| $X_3$ | 0.252 | 0.546 | 0.372 | 0.132 |
| $X_4$ | 0.152 | 0.022 | 0.014 | 0.028 |
| $X_5$ | 0.085 | 0 | 0 | 0.01 |
| $X_6$ | 0.233 | 0.316 | 0.800 | 0.662 |
| $X_7$ | 0.281 | 0.812 | 0.928 | 0.71 |
| $X_8$ | 0.228 | 0.388 | 0.768 | 0.382 |
| $X_9$ | 0.280 | 0.786 | 0.928 | 0.728 |
| $X_{10}$ | 0.032 | 0 | 0.024 | 0.006 |

One unexpected finding was the very low true positive rate for certain variables. For the Bonferroni-adjusted univariable analyses, we would expect each variable to be detected in 39-40% of simulations, based on power to detect effects in training and validation sets. However, the proportion of time a variable was chosen (binary case) ranged from 0-97% (see Table 3). These results were attributable to confounding within $X_T$. Confounding of the relationship between a predictor $X$ and an outcome $Y$ occurs when a third factor is associated with both $X$ and $Y$. Here, there were small to moderate correlations between the members of $X_T$ (see Additional File 4). Therefore, when variables were analyzed separately, the regression model was misspecified due to confounding. As shown in Table 3, the estimated effects from univariable models (i.e. our UNIV-BFN models) relate to the proportion of times a variable is identified. Variables with estimated effects >0.22 (larger than the truth) were more likely to be selected, while variables with estimated effects <0.22 (smaller than the truth; $X_4$, $X_5$, $X_{10}$) were almost never identified. Unfortunately, even methods like the lasso which simultaneously consider all variables did not provide substantial improvements in detection rates of rarely selected variables, nor did using a multivariable (lasso) model in the validation step in place of the univariable regressions.

## 2.3 Application to PA prostate cancer cases

We assessed associations between the census variables and the outcome of aggressive Prostate Cancer (PCa) PCa using HCLST-CORR-SGL, the best-performing method for binary outcomes. After applying the hierarchical clustering algorithm with a threshold of 0.8, 10,888 of the 14,663 variables were grouped with at least one other variable. Of the 6,865 clusters identified, 3,090 had two or more variables (max 244), and 3765 clusters contained only one variable. HCLST-CORR-SGL identified nineteen census variables in thirteen clusters as predictors of aggressive disease (Additional file 5), as well as year of diagnosis and patient age. One variable found in the NWAS study was identified by this approach, and two variables/clusters in the NWAS were highly related to variables identified by HCLST-CORR-SGL. These overlapping variables are described in Table 4. We note that the results from HCLST-CORR-SGL were sparse within clusters; nevertheless, the algorithm does not force selection of a single representative variable from each cluster. Therefore, some highly related variables were chosen (e.g. PCT_SF3_PCT065I007 and PCT_SF3_PCT065A007).

Table 4. Results from full data: Variables identified as associated with PCa aggressiveness by both HCLST-CORR-SGL and NWAS

| SGL variable(s) | Domain: Description | NWAS variable | Correlation(s) |
|---|---|---|---|
| PCT_SF3_PCT050102 | Poverty: % Ratio of Income to Poverty level for persons aged 45-54 under 0.50 | PCT_SF3_PCT050102 | 1.0 |
| PCT_SF3_HCT005092 | Housing/Income: %Renter-occupied housing units built 1939-1949 with householder aged 25-34 | PCT_SF3_HCT015042 | 0.912 |
| PCT_SF3_PCT065I007 | Employment/Transportation: % White Only (non-Hispanic) Worker taking public transportation (trolley or streetcar) to work | PCT_SF3_P030007 | 0.935 |
| PCT_SF3_PCT065A007 | % White Only Worker taking public transportation (trolley or streetcar) to work | | 0.935 |

## 3. DISCUSSION AND CONCLUSIONS

In simulation studies, we found that methods using hierarchical clustering combined with sparse group lasso (HCLST-CORR-SGL) performed the best at identifying variables with true associations (or their surrogates), while providing control of false positive results. This conclusion is based on the method's F2 scores in simulated data, a combined measure of accuracy which gives greater weight to the method's sensitivity. HCLST-CORR-SGL used clustering to directly address the complex correlation structure of the data, which may have led to improvements in the ability of penalized regression to detect true positive variables. We showed that the simpler threshold-based approach was sufficient to define meaningful clusters. However, this approach, like the others we assessed, did not solve the low detection rates of variables subject to confounding.

We applied the HCLST-CORR-SGL to our full dataset and compared findings from these machine learning methods to our previously published NWAS, under the assumption that variables that replicated across methods were more likely to represent true findings. We note that in our simulation study, outcomes were generated completely independently, while the observed outcomes in the full dataset likely had spatial effects which were not accounted for in the machine learning approaches applied here. Nevertheless, HCLST-CORR-SGL did independently replicate three out of seventeen NWAS variables (i.e. the identification of the same or a highly correlated variable) which did take into account potential spatial effects.

Previous studies of social-environmental factors often selected census variables *a priori*. These studies showed that single variables representing single domains (e.g. % living below poverty) were associated with advanced prostate cancer and cancer more broadly (32). Interestingly, the NWAS and machine learners consistently identified more complex variables that combined domains related to race, age, and poverty with household or renter status. Thus, findings from these empirical methods could serve to be hypothesis-generating, suggesting interactions among domains that are often considered individually in current social-environmental studies. For example, the top hit from the previous NWAS (PCT_SF3_P030007) had a correlation of 0.93 with two variables identified by HCLST-CORR-SGL (PCT_SF3_PCT065I007 and PCT_SF3_PCT065A007). All three are markers of employment and transportation, a combination of two different domains.

This study has several limitations. First, although we assessed several popular machine-learning algorithms for variable selection, there are many other approaches. We considered principal components regression, as it is commonly used with highly collinear data, but ultimately did not include it because the results were difficult to interpret and required arbitrary thresholds. Other popular machine learning approaches that use a "black box" algorithm for prediction (e.g. neural networks) were not readily useable for variable selection and therefore were not included. Most Support Vector Machines (SVM) based methods are not readily used for variable selection; we considered a sparse SVM method, but found that it was computationally infeasible to implement. (33) We also did not evaluate predictive accuracy of the various methods, as it was not of primary interest. Further, we intentionally designed a situation where variables had small effects compared to the random error, (10) so even a perfect method would have relatively low predictive ability. In real-world cases, the social-environmental variables would be combined with patient-level variables, giving the models much better predictive accuracy; we did not do so here to isolate the effect of method choice on selection of neighborhood predictors. Finally, for computational tractability, the size of the dataset used in simulations was limited to 1,000 variables and 2,000 subjects, much smaller than the full dataset upon which this study is based. In future studies, we will assess the impact of spatial relationships and the rate of true associations on the method's performance. We will also consider cases where the data-generating model is non-linear, includes interactions, and uses patient-level predictors.

In this era of Big Data and Precision Medicine, (34, 35) the importance of neighborhood and other environmental data will continue to grow. Given the complex structure and high dimensionality of environmental data, researchers should continue to develop machine learning approaches for this area. For complex diseases like cancer, the analysis of multilevel, mixed feature datasets (including environmental, biological, and behavioral features) will likely be needed to inform health disparities, disease prevention and clinical care, motivating the development of new analytical approaches.

**Acknowledgements**

We would like to thank Dr. Karthik Devarajan for his expert consultation on machine learning methods and computational approaches and Kristen Sorice for her assistance with this manuscript submission. Prostate cancer case data were supplied by the Pennsylvania Department of Health who disclaims responsibility for any analyses, interpretations, or conclusions.

**Competing Interests**

None.

**Funding**

This work is supported by grants from the American Cancer Society IRG-92-027-20 and MRSG CPHPS-130319 to SML. This research was also funded in part through the NIH/NCI Cancer Center Support Grant P30 CA00692 and NIHU54 CA221705.  The content is solely the responsibility of the authors and does not necessarily represent the official views of the National Institutes of Health

Additional Files:

1. Additional file 1

   TIFF (.tiff) file

   Title: Dendrogram for correlation between variables

   Description: Dendrogram showing the relationships between the 1,000 elements of the covariate matrix X. The horizontal red line represents a correlation of 0.8.

2. Additional file 2

   TIFF (.tiff) file

   Title: Distribution of 10 variables associated with simulated outcomes

   Description: Histograms showing the distributions of $X_1$-$X_{10}$, the elements of $X$ used to simulate the outcomes Y.

3. Additional file 3

   Excel (.xlsx) file

   Title: Detection rates for each variable

   Description: Table showing the proportion of simulations where each varaible was selected (by each method)

4. Additional file 4

   TIFF (.tiff) file

   Title: Correlation structure of 10 variables

   Description: Correlations structure of $X_1$-$X_{10}$, the elements of X used to simulate the outcomes Y. Blue represents a positive correlation and red a negative correlation, with darker colors indicating a stronger relationship.

5. Additional file 5

   Excel (.xlsx) file

   Title: Full data results

   Description: Table containing all finidings when the HCLST-CORR-SGL method was applied to the full PA PCa dataset (binary outcome: diagnosis with aggressive PCa)